\documentclass[12pt]{iopart}
\usepackage[dvips]{epsfig}
\usepackage[latin1]{inputenc}

\usepackage{color}

\begin{document}

\title[Outflow Dynamics in Modeling Oligopoly Markets]{Outflow Dynamics in Modeling Oligopoly Markets: The Case of the Mobile Telecommunications Market in Poland}

\author{Katarzyna Sznajd-Weron$^1$, Rafa{\l} Weron$^2$ and Maja W{\l}oszczowska$^3$}
\address{$^1$ Institute of Theoretical Physics, Wroc{\l}aw University, 
pl. Maxa Borna 9, 50-204 Wroc{\l}aw, Poland}
\address{$^2$ Hugo Steinhaus Center for Stochastic Methods, 
Wroc{\l}aw University of Technology, Wyb. Wyspia\'nskiego 27, 50-370 Wroc{\l}aw, Poland}
\address{$^3$ Institute of Mathematics and Computer Science,
Wroc{\l}aw University of Technology, Wyb. Wyspia\'nskiego 27, 50-370 Wroc{\l}aw, Poland}
\ead{kweron@ift.uni.wroc.pl, rafal.weron@pwr.wroc.pl and 126582@student.pwr.wroc.pl}

\begin{abstract}
In this paper we introduce two models of opinion dynamics in oligopoly markets and apply them to a situation, where a new entrant challenges two incumbents of the same size. The models differ in the way the two forces influencing consumer choice -- (local) social interactions and (global) advertising -- interact. We study the general behavior of the models using the Mean Field Approach and Monte Carlo simulations and calibrate the models to data from the Polish telecommunications market. For one of the models criticality is observed -- below a certain critical level of advertising the market approaches a lock-in situation, where one market leader dominates the market and all other brands disappear. Interestingly, for both models the best fits to real data are obtained for conformity level $p \in (0.3,0.4)$. This agrees very well with the conformity level found by Solomon Asch in his famous social experiment.


\end{abstract}

\section{Introduction}

Economic and social studies suggest that the decision of a consumer to purchase a particular product depends not only on price and product quality, but also on social effects and advertising exposure \cite{erd:kea:96,jan:jag:01}. On one hand, the social validation phenomenon is a powerful motivation for our actions. People listen to their family and friends' experiences of network operators, price plans and brands of handsets \cite{tur:lee:yin:00,lee:cha:06}. The tariffs themselves add to this clustering effect by giving preferences (including lower prices) to connections within a network or even within a group of friends. 

On the other hand, the willingness to pay for a product is increased by advertisements \cite{bec:mur:93}. Advertising is used to reinforce the brand image as it is used as an indicator of quality and makes services such as telecommunications more tangible to consumers. Indeed, Turnbull et al. \cite{tur:lee:yin:00} found advertising to be quite an important source of information when purchasing a mobile phone, but not as important as family and friends. At the same time, mobile telephony is the most heavily advertised business in many countries \cite{soh:cho:01}. It seems that the operators are aware of the power of advertising. And since they cannot influence social relations directly, they are doing the second best thing -- advertising. 

But what affects consumer decisions more? Is it social influence or advertising? The aim of this paper is to build a model of the mobile telecommunications market in Poland, where three major operators (\emph{Plus}, \emph{Era} and \emph{Orange}) fight an on-going battle for domination. To tackle this problem we introduce two models of opinion dynamics in oligopoly markets, which differ in the way the two forces influencing consumer choice -- (local) social interactions and (global) advertising -- interact. We study the general behavior of the models: steady states in terms of market shares for the entire space of parameters, as well as calibrate them to real data. Hopefully, the obtained results will add to the understanding of oligopoly market behavior and help in answering the above question.
 
The paper is structured as follows. In the next Section we provide a short overview of the mobile phone market in Poland. In Section \ref{sec:TheModel} we discuss the factors influencing our choice of the mobile operator and introduce two opinion dynamics models. We conclude the Section by presenting Mean Field equations for both models.
In Section \ref{sec:TheResults} we present the obtained results both in terms of Monte Carlo simulations and Mean Field approximations. Finally, in Section \ref{sec:Conclusions} we conclude and discuss possible follow-up research.

\section{The Market}
\label{sec:TheMarket}

\subsection{The Mobile Telecommunications Market in Poland}

Mobile telephony in Poland dates back to 1992, when PTK (brand \emph{Centertel}) started offering services in the analog NMT450i system. The first mobile phones were the `bricks' (Nokia Cityman 450) and the `radiators' (Motorola 2000, Talkman 900). Their price exceeded the average annual salary at that time. In the early 1990s mobile phones were the synonym of success in the rapidly developing Polish economy. Second generation services (GSM 900 MHz) were offered in 1996 by two new operators -- PTC (brand \emph{Era}) and Polkomtel (brand \emph{Plus}). Only then, through competition and the policy of subsidizing mobile phones have they become more popular. Centertel started loosing clients due to worser quality of transmission and bulkier handsets and in 1998 decided to launch a digital system of its own -- \emph{Idea Centertel} (rebranded to \emph{Orange} in 2005) working in the GSM 1800 MHz system. In late 1998 all three operators made a next major step forward and introduced prepaid services. This move eventually led to mobile phones becoming everyday appliances.

In 2000 the operators received concessions for new frequencies and since then all three have used both the 900 and 1800 MHz systems. This was a great opportunity especially for \emph{Idea Centertel}, which has been using the shorter range, city-concentrated GSM 1800 MHz system and had three times fewer clients than the two competitors. The next major technological change -- introduction of the third generation UMTS telephony in 2004 -- did not have a noticeable impact on the market. Even today there are not too many users of this system. In March 2007 a fourth player entered the market -- P4 (brand \emph{Play}), but by the end of 2007 its market share was still negligible ($2.05\%$ in terms of the number of users and $1.28\%$ in terms of net revenues).
 
During the last 15 years the number of mobile phone users has been steadily increasing. In 2007 the number of sold SIM cards exceeded 40 million yielding a market penetration of 108.6\%; the penetration by active SIM cards was estimated at 90.9\% \cite{uke:08}. Compared to other European countries this is not a spectacular result.

The mobile phone market is a typical example of an oligopoly. Entry barriers (including infrastructure and concessions) are extremely high. As of end 2007 there were four operators in the Polish market \cite{uke:08}:
\begin{itemize}
\item Polkomtel S.A. (brand: \emph{Plus}, prepaid brands: \emph{Simplus} and \emph{Sami Swoi})
\item Polska Telefonia Cyfrowa Sp. z o.o. (brand: \emph{Era}, prepaid brands: \emph{Tak Tak} and \emph{Heyah})
\item Polska Telefonia Kom\'orkowa Centertel Sp. z o.o. (brand \emph{Orange}; includes prepaid services)
\item P4 Sp. z o.o. (brand \emph{Play}; includes prepaid services)
\end{itemize}
There were also a few Mobile Virtual Network Operators (MVNO) -- brands \emph{mBank mobile}, \emph{myAvon}, \emph{WPmobi}, \emph{Simfonia} and \emph{Ezo} -- but their market share was negligible ($0.12\%$ in terms of the number of users and $0.02\%$ in terms of net revenues as of end 2007). 

\begin{figure}[tbp]
\centering
\includegraphics[width=12cm]{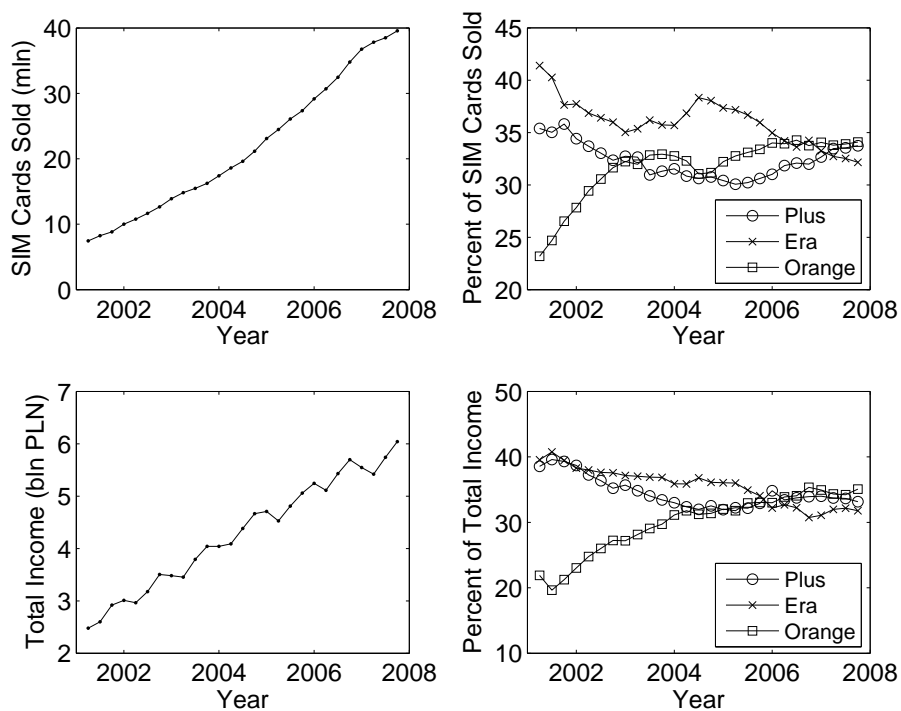}
\caption{\emph{Left panels:} The cumulative number of SIM cards sold and the total income (i.e., jointly for all three major market players) from the first quarter of 2001 till the 3rd quarter of 2007. A steady upward trend can be observed in both cases, with noticeable seasonality in income figures. \emph{Right panels:} Market structure in terms of SIM cards sold and income. The turmoil around PTC led to a depreciation of the company's leading brand \emph{Era} and, consequently, the company's financial standing.}
\label{fig:SIM_Income}
\end{figure}

In this paper we study the period from the end of year 2000, when the technological constraints of the three major operators were leveled out, till the third quarter of 2007, see Figure \ref{fig:SIM_Income}. We will only consider these three players (and generally denote them later in the text by their main brands -- \emph{Plus}, \emph{Era} and \emph{Orange}) as the remaining operators have not played a visible role in the studied period. 

For a long time these three players have fought an on-going battle for domination in the Polish mobile phone market. But despite the competition, they have sustained relatively high prices for their services and tempted new clients with cheap handsets instead (similar strategies have been observed in other markets as well \cite{cho:lee:chu:01}). Only when the market has saturated (2004-2005), the operators have started attracting the less rich clients. First, by lowering the prices for prepaid connections, next by lowering the prices for subscription (postpaid) services. The latter move has reversed the dominating trend -- instead of offering cheap handsets the operators have started luring clients with cheap connections. 

Apart from the service-handset price competition, the operators have always fought a war on the billboards and in radio, TV and internet commercials. In fact, in the last ten years the three major mobile market players have been the most heavily advertised companies in Poland. Nearly every week they make the top ten list, if not the top three. Mobile phone ads are visible everywhere and it is hard to imagine a new market entry without an intensive advertising campaign.

\subsection{The turmoil around PTC}
\label{ssec:TurmoilERA}

Writing about the Polish mobile telecommunications market we have to mention the controversial events related to the dispute over 48\% of shares of PTC (owner of the brand \emph{Era}). Four companies were involved in this: Elektrim, Elektrim Telekomunikacja, Deutsche Telekom and Vivendi. The dispute started in 1999, when Elektrim bought 15.8\% of shares from other shareholders (Kulczyk Holding and BRE Bank). At the same time, another shareholder -- Deutsche Telekom (DT) -- was convinced that it had the preemtion right to those shares and in December 2000 filed a case in the Arbitrage Court in Vienna. Legal actions ended in November 2004. In the meantime, the shareholder structure changed and the sentence did not hurt Electrim but a French company Vivendi. In 1999 Vivendi bought 49\% of shares of Elektrim Telekomunikacja (ET), a company to which Elektrim transferred its PTC shares in 2000. In 2001 Vivendi bought another 2\% of shares of ET gaining control not only over ET but also over PTC. However, the Arbitrage Court ruled that the transfer of shares from Elektrim to ET was illegal. The Provincial Court in Warsaw (S\c{a}d Rejonowy w Warszawie) recognized this decision and 48\% of PTC shares returned to Elektrim. And then all the fuss began.

In February 2005 the Provincial Court in Warsaw made changes in the National Company Register (Krajowy Rejestr S\c{a}dowy, KRS), including a new board of directors and a new supervisory board appointed by Elektrim in cooperation with DT. ET and Vivendi did not accept this, which resulted in a blockade of PTC's headquarters by the former management. Vivendi also filed a complaint to the Polish government, backed by a bilateral agreement between France and Poland on observance of international investments. After some time the new board of directors gained control over PTC's offices and Elektrim started negotiations with DT and Vivendi on the sale of PTC's shares. However, in August 2005 the roles changed: the District Court in Warsaw (S\c{a}d Okr\c{e}gowy w Warszawie; playing the role of the court of appeal for the Provincial Court) reversed the decision of the lower instance and sent the case back to the Provincial Court in Warsaw. In November 2005 the Provincial Court made changes in KRS in favor of ET. This time the Elektrim and DT appointed board blocked PTC's headquarters. A number of managing directors resigned. PTC started to have liquidity problems.

\begin{figure}[tbp]
\centering
\includegraphics[width=12cm]{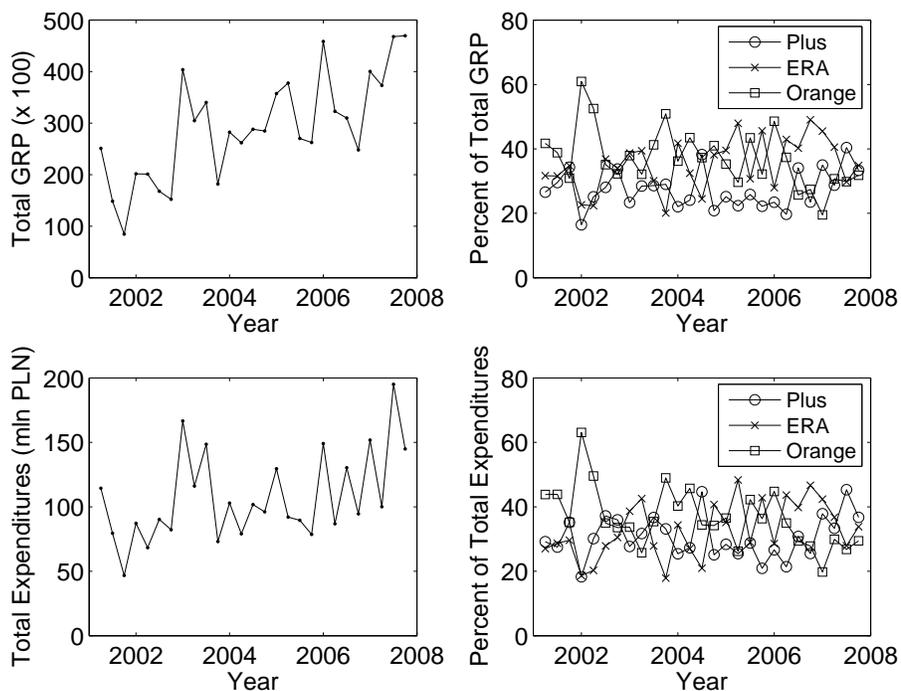}
\caption{\emph{Left panels:} Quarterly Gross Rating Points (GRP) and advertising expenditures (jointly for all three major market players) for the same period as in Figure \protect\ref{fig:SIM_Income}. The processes are not identical, however, contrary to some reports \protect\cite{dub:hit:man:05} they are not that different (the correlation for individual companies ranges from $\rho=0.92$ to $0.95$). \emph{Right panels:} Market structure in terms of GRP and advertising expenditures.}
\label{fig:GRP_Invest}
\end{figure}

It seems that the case was finally resolved in March 2006, when the Warsaw Court of Appeal (S\c{a}d Apelacyjny w Warszawie) upheld the sentence of the Arbitrage Court in Vienna from November 2004. The situation of PTC stabilized, although courts of different instances were still processing the case for two more years. During this turmoil \emph{Era} lost its pole position in the mobile phone market in Poland (Fig. \ref{fig:SIM_Income}). This happened despite the fact that advertising remained high, both in terms of advertising expenditures and Gross Rating Points (Fig. \ref{fig:GRP_Invest}). Some authors \cite{dub:hit:man:05} consider the Gross Rating Points (GRP), i.e. the percentage number of targeted viewers contacted times the average number of contacts per targeted viewer, to be much more representative of the actual advertising campaign. In our case, however, the advertising expenditures and GRP numbers yield a very similar picture. Also the correlation between these two variables is very high and for individual companies ranges from $\rho=0.92$ to $0.95$ in the studied period.

A March 2006 survey performed by PBS (www.pbsdga.pl), a Polish market research company, sheds some light on the discord between PTC's market share and advertising expenditures. Nearly 40\% of respondents (and over 55\% of corporate respondents) declare that news of an uncertain future of the company would make them change the operator. Also over 30\% of respondents would change the operator if serious charges were brought against the company's management.

\section{The Model}
\label{sec:TheModel}

\subsection{Factors Influencing Our Choice of the Mobile Operator}

We generally prefer to make decisions consciously. In an attempt to make the best choice possible we study the operators' offers: special deals, price plans, offered handsets, loyalty programs, network coverage and quality of transmission. Especially subscription clients spend long hours on studying and weighting these factors. Quite often, however, the offers are very similar when we take all factors into account and it is very hard to make a decision based only on simple statistical comparisons. 

Economic and social studies suggest that the decision of a consumer to purchase a particular product depends not only on price and product quality, but also on social effects and advertising exposure \cite{erd:kea:96,jan:jag:01}. Moreover, links in a social network and the stream of advertisements may lead to common actions of large groups of consumers and induce correlation between consumers' decisions \cite{gro:06}.

On one hand, the social validation phenomenon is a powerful motivation for our actions. People listen to their family and friends' experiences of network operators, price plans and brands of handsets. Turnbull et al. \cite{tur:lee:yin:00} found family and friends to be the main source of information used by UK consumers when purchasing a mobile phone. A similar effect was observed in the Thai market \cite{lee:cha:06}. Note that tariffs add to this clustering effect by giving preferences (including lower prices) to connections within a network or even within a group of friends. 

On the other hand, according to Becker and Murphy \cite{bec:mur:93} the willingness to pay for a product is increased by advertisements. This implies that when a positive news arrive at a consumer, they raise his/her perception of utility of that brand. Advertising is used to reinforce the brand image as it is used as an indicator of quality and makes services such as telecommunications more tangible to consumers. Indeed, Turnbull et al. \cite{tur:lee:yin:00} found advertising to be quite an important source of information when purchasing a mobile phone in the UK (but not as important as family and friends). At the same time, mobile telephony is the most heavily advertised business in many countries, see Section \ref{sec:TheMarket} and Ref. \cite{soh:cho:01}. It seems that the operators are aware of the power of advertising. And since they cannot influence social relations directly, they are doing the second best thing -- advertising.

\subsection{Building the Model}
\label{ssec:BuildingTheModel}

From the previous Section we know that social influence plays a major role in selecting a mobile phone company. But can we be more precise and say which type of influence is more and which is less important for the decision making process? 

In the early 1950s Solomon Asch reported an ingenious series of experiments on social influence (for a fascinating review see Levine \cite{lev:99}). Asch used several measures of social influence but here we recall only two of his results \cite{asc:55}, which help us answer the question `Which aspect of the influence of a majority is more important -- the size of the majority or its unanimity?':
\begin{enumerate} 
\item
Asch found that minority participants confronting a unanimous majority were often capable of resisting social influence. When a subject was confronted with only a single individual who contradicted his answers, he was swayed little: he continued to answer independently and correctly in nearly all trials. When the opposition was increased to two, the pressure became substantial: minority subjects now accepted the wrong answer 13.6 \%  of the time. Under the pressure of a majority of three, the subjects' errors jumped to 32 \%. But further increases in the size of the majority apparently did not increase the weight of the pressure substantially. Clearly the size of the opposition is important only up to a point.
\item
Asch also found that he could increase independence dramatically -- the presence of a social supporter reduced the total number of yielding responses from 32\% to 5.5\% \cite{asc:55}. The power of social support was further demonstrated in a study showing that participants were far more independent when they were opposed by a seven-person majority and had a partner than when they were opposed by a three-person majority and did not have a partner \cite{lev:99}.
\end{enumerate}

A number of later experiments showed that an individual who breaks the unanimity principle reduces social pressure of the group dramatically \cite{mye:06}. This observation was recently expressed in a simple one dimensional USDF (`United we Stand, Divided we Fall') model of opinion formation \cite{s-w:szn:00}. The model was later renamed `the Sznajd model' by Stauffer et al. \cite{sta:sou:oli:00} and generalized on a two dimensional square lattice. 
The crucial difference between the Sznajd model and other Ising-type models of opinion dynamics \cite{gal:86,gal:90,kra:red:03} is that information flows outward from the center nodes to the surrounding neighborhood (so called outflow dynamics \cite{s-w:kru:06}) and not the other way around. Ising-type models with outflow dynamics have been successfully applied in marketing, finance and politics; for reviews see \cite{sta:02,sch:02,for:sta:05,s-w:05,cas:for:lor:07}.

The aim of this paper is to build a model of the mobile telecommunications market in Poland, where three major operators (\emph{Plus}, \emph{Era} and \emph{Orange}) fight an on-going battle for domination. The market is represented by a two dimensional $L \times L$ lattice with periodic boundary conditions. Each site of the lattice is occupied independently by an individual (a customer), who is characterized by a variable $S_i=1,2,3, \; i=1,...,L^2$, that represents his/her mobile operator. In each time step ($\Delta\tau$) one of the $L^2$ customers is selected randomly (random sequential updating) and then together with his/her three neighbors forms a $2\times 2$ panel. We measure the time so that the speed of all processes remains constant when $L \rightarrow \infty$, and thus one update takes time $\Delta\tau=1/L^2$.

Two forces influencing consumer choice are considered: (local) social interactions and (global) advertising. Some empirical studies \cite{tur:lee:yin:00} suggest that of the two forces social interactions play a greater role. Following the unanimity principle discovered by social scientists we assume that only an unanimous panel (all four customers in the panel use the same mobile network) persuades all eight of its nearest neighbors to switch the operator to the one favored by the panel. In lack of unanimity the eight neighboring individuals make their choice based on advertisements: independently with probability $1-p$ each of them switches the operator. The choice of the operator is determined by the relative advertising level (a kind of an external field \cite{sch:03,s-w:wer:03,can:maz:08}). With probability $h_1$ he/she chooses company $1$ (say, \emph{Plus}), with probability $h_2$ company $2$ (\emph{Era}) and with probability $h_3=1-h_1-h_2$ company $3$ (\emph{Orange}). Obviously, the customer may `switch' to the same operator, i.e., stay with the original one. In this model conformity (social influence) comes in first, then advertising affects the customers. Hence, we will call the model -- Conformity First (CF).

A different idea of incorporating advertising in a duopoly setting was proposed by the social psychologist Andrzej Nowak at the GIACS summer school `Applications of Complex Systems to Social Sciences' \cite{wol:sta:kul:07}. He suggested to generalize the USDF model by flipping each up opinion down ($S_i=1 \rightarrow S_i=-1$) with probability $\gamma$ and each down opinion up ($S_i=-1 \rightarrow S_i=1$) with probability $\beta$ as a result of global effects like advertising through mass media. The traditional conformity rule was applied to each of the neighbors (of the panel) independently with probability $\alpha$. In this model (called `Nowak-Sznajd' by Wo{\l}oszyn et al. \cite{wol:sta:kul:07}) conformity influences individuals parallel to advertising. We borrow Nowak's idea to construct our second model and call it CAP (Conformity and Advertising Parallel). 

In the CAP model, for each of the the eight nearest neighbors of the selected panel we independently apply either the social (with probability $p$) or advertising (with probability $1-p$) forces. All neighbors selected to be influenced by social interactions switch the operator to the one favored by the panel, but only if the panel is unanimous. In lack of unanimity the selected neighbors do not perform any action. The remaining neighbors (selected to be influenced by advertising) are subdued to the same rules as in the CF model. Namely, the choice of the operator is determined by the relative advertising level: with probability $h_1$ the individual chooses company $1$, with probability $h_2$ company $2$ and with probability $h_3=1-h_1-h_2$ company $3$. Note, that in contrast to the CF model, in the CAP model an unanimous panel does not guarantee conformity.

\subsection{Mean Field Approach}
\label{ssec:MFA}

Let us denote by $N_1(t)$, $N_2(t)$ and $N_3(t)$ the number at time $t$ of \emph{Plus}'s, \emph{Era}'s and \emph{Orange}'s customers, respectively. Further, define by $c_1(t)=\frac{N_1(t)}{L^2}$, $c_2(t)=\frac{N_2(t)}{L^2}$ and $c_3(t)=\frac{N_3(t)}{L^2}$ the respective market shares (concentrations). Of course, the normalization condition is fulfilled, i.e. $\forall t \hspace{0.3cm} c_1(t)+c_2(t)+c_3(t)=1$. We measure the time so that the speed of all processes remains constant when $L \rightarrow \infty$, and thus one update takes time $\Delta\tau=1/L^2$. Each basic time step $\Delta t=1$ consists of $L^2$ updatings, hence the balance (evolution) equation takes the form:
\begin{equation}
N_s(t+1) - N_s(t) = L^2 (\mbox{incremental changes} - \mbox{decremental changes}),
\end{equation}
where $s=1,2,3$.

The incremental and decremental changes are (slightly) different in the two models. For instance, in the CAP model the number $N_1(t)$ of \emph{Plus}'s (operator \#1) clients can increase in time due to the following events:
\begin{itemize}
\item
an \emph{Era} client (operator \#2) follows \emph{Plus}'s advertising -- the probability of such an event is equal to $(1-p)h_1c_2(t)$;
\item
an \emph{Orange} client (operator \#3) follows \emph{Plus}'s advertising -- the probability of such an event is equal to $(1-p)h_1c_3(t)$; 
\item
an \emph{Era} client subdues to the unanimous panel of operator \#1 -- the probability of such an event is equal to $pc_1(t)^4c_2(t)$; 
\item
an \emph{Orange} client subdues to the unanimous panel of operator \#1 -- the probability of such an event is equal to $pc_1(t)^4c_3(t)$.
\end{itemize}
Similarly, the number $N_1(t)$ of \emph{Plus}'s clients can decrease in time due to the following events:
\begin{itemize}
\item
a \emph{Plus} client follows \emph{Era}'s advertising -- the probability of such an event is equal to $(1-p)h_2c_1(t)$;
\item
a \emph{Plus} client follows \emph{Orange}'s advertising -- the probability of such an event is equal to $(1-p)h_3c_1(t)$;
\item
a \emph{Plus} client subdues to the unanimous panel of operator \#2 -- the probability of such an event is equal to $pc_2(t)^4c_1(t)$;
\item
a \emph{Plus} client subdues to the unanimous panel of operator \#3 -- the probability of such an event is equal to $pc_3(t)^4c_1(t)$.
\end{itemize}

Now, we can put down the balance equation for the first operator:
\begin{eqnarray}
N_1(t+1) - N_1(t)& = & L^2 [(1-p)h_1c_2(t)+ (1-p)h_1c_3(t) \nonumber \\ 
& + & pc_1(t)^4c_2(t)+ pc_1(t)^4c_3(t) \nonumber \\ 
& - & (1-p)h_2c_1(t) - (1-p)h_3c_1(t) \nonumber \\
& - & pc_2(t)^4c_1(t) - pc_3(t)^4c_1(t)].
\end{eqnarray} 
Dividing both sides of the equation by $L^2$ and denoting $c_s(t)$ by $c_s$ and $c_s(t+1)$ by $c'_s$, for $s=1,2,3$, after simple algebraic transformations we obtain:
\begin{eqnarray}
c'_1 - c_1& = &(1-p)h_1(c_2+c_3)-c_1(1-p)(h_2+h_3)\nonumber \\
& + & pc_1c_2(c_1^3-c_2^3)+pc_1c_3(c_1^3-c_3^3).
\end{eqnarray} 
Applying the normalization conditions ($c_2+c_3=1-c_1$ and $h_2+h_3=1-h_1$) we obtain the complete set of evolution equations for the CAP model:
\begin{eqnarray}
c'_1 - c_1 & = &(1-p)(h_1-c_1) + pc_1 \left[c_2(c_1^3-c_2^3)+c_3(c_1^3-c_3^3) \right], \nonumber\\
c'_2 - c_2 & = &(1-p)(h_2-c_2) + pc_2 \left[c_1(c_2^3-c_1^3)+c_3(c_2^3-c_3^3) \right], \nonumber\\
c'_3 - c_3 & = &(1-p)(h_3-c_3) + pc_3 \left[c_1(c_3^3-c_1^3)+c_2(c_3^3-c_2^3) \right]. \label{eqn:MFA:CAP}
\end{eqnarray} 
Analogously, we can derive the evolution equations for the CF model:
\begin{eqnarray}
c'_1 - c_1 & = &(1-p)(h_1-c_1) + c_1 \left[c_2(c_1^3-c_2^3)+c_3(c_1^3-c_3^3) \right], \nonumber\\
c'_2 - c_2 & = &(1-p)(h_2-c_2) + c_2 \left[c_1(c_2^3-c_1^3)+c_3(c_2^3-c_3^3) \right], \nonumber\\
c'_3 - c_3 & = &(1-p)(h_3-c_3) + c_3 \left[c_1(c_3^3-c_1^3)+c_2(c_3^3-c_2^3) \right]. \label{eqn:MFA:CF}
\end{eqnarray} 
Note, that for $p=1$ the CAP and CF evolution equations are identical and that the differences increase with decreasing $p$. Note also, that the sets of evolution equations (\ref{eqn:MFA:CAP}) and (\ref{eqn:MFA:CF}) can be easily solved numerically, see the discussion in Section \ref{sec:TheResults}.

\subsection{The Model of the Polish Telecommunications Market (2000-2007)}
\label{ssec:PolishMarket}

Recall from Section \ref{sec:TheMarket} that back in year 2000 (beginning of the study period) three mobile operators were offering their services in the Polish market: a `new entrant' (\emph{Orange}; until 2005 under the brand \emph{Idea Centertel}) was challenging two incumbents (\emph{Plus} and \emph{Era}). The market shares of the two incumbents were nearly the same in terms of income (ca. 40\% each; see the bottom right panel in Figure \ref{fig:SIM_Income}), but not in terms of the number of SIM cards sold. Following W{\l}oszczowska \cite{wlo:08}, in this study we use income as the measure of market share. There are two major reasons for this. First, the published data concern the number of SIM cards sold, not the actual number of users of a given network. Some of the cards are not active and some clients are using more than one SIM card (private, business, internet). Second, different categories of clients use the phones with different intensities. This could be accounted for by assigning more lattice sites to some clients, but for the sake of parsimony we have decided not to execute this option. Consequently, in our model one lattice node represents an average (in terms of income) client.

The market state where a new entrant challenges two incumbents of the same size allows us to make further simplifications. We limit the rich parameter space ($p, h_s, c_s(0)$) to the situation where the initial concentrations of clients and the levels of advertisement of \emph{Plus} and \emph{Era} are equal, i.e. $c_1(0)=c_2(0)\equiv c_0$ and $h_1=h_2\equiv h$, respectively.

\section{The Results}
\label{sec:TheResults}

For the sets of rules governing the behavior of mobile phone users (see Section \ref{ssec:BuildingTheModel}), we study the system via Monte Carlo (MC) simulations and compare the final steady states with those obtained from the Mean Field Approach (MFA), see equations (\ref{eqn:MFA:CAP}) and (\ref{eqn:MFA:CF}).

\subsection{Monte Carlo Simulations}

We present results for lattice size $100 \times 100$, although we have performed simulations for other $L$'s. In all simulations we took initially $c_0$ of \emph{Plus} customers, $c_0$ of \emph{Era} customers and $1-2c_0$ of \emph{Orange} customers randomly distributed on the lattice. The level of advertisement $h$ is the same for \emph{Plus} and \emph{Era} and constant throughout the simulations. For \emph{Orange} the level is equal to $1-2h$. 

\begin{figure}[tbp]
\centering
\includegraphics[width=13cm]{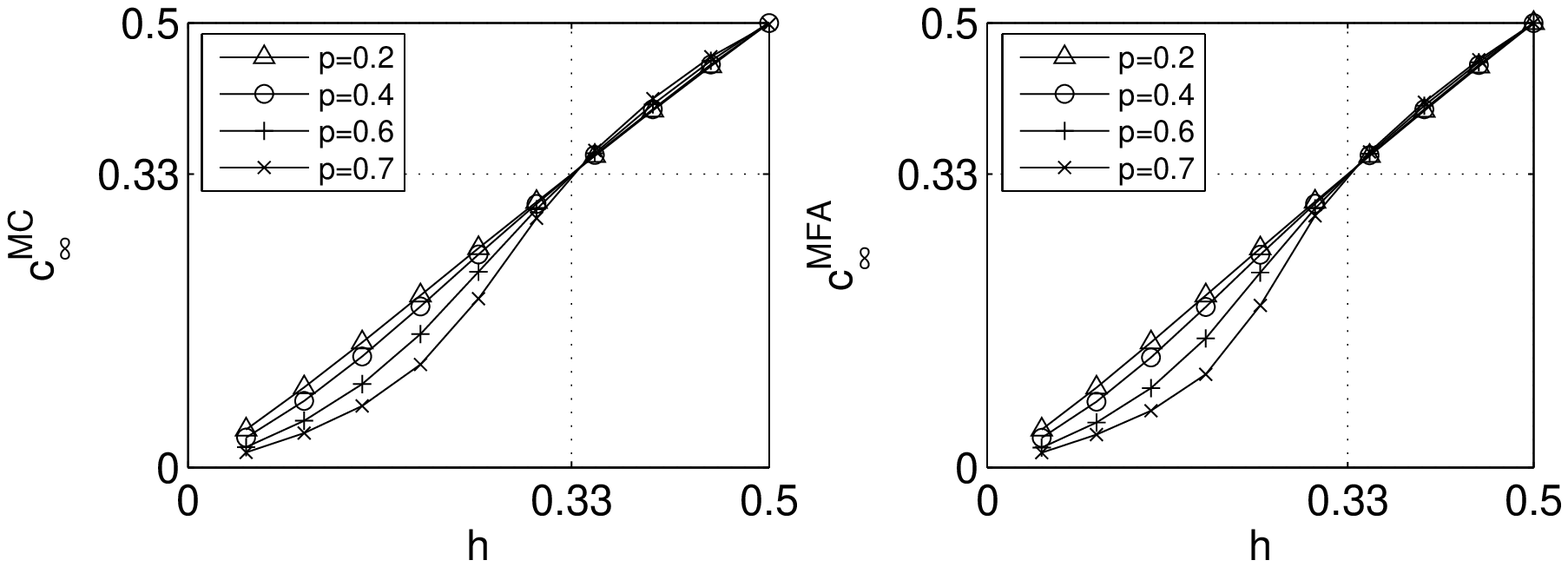}
\caption{MC and MFA results for the CAP model and $p\le 0.7$}
\label{fig:fig_c_inf1_n}
\end{figure}

\begin{figure}[tbp]
\centering
\includegraphics[width=12cm]{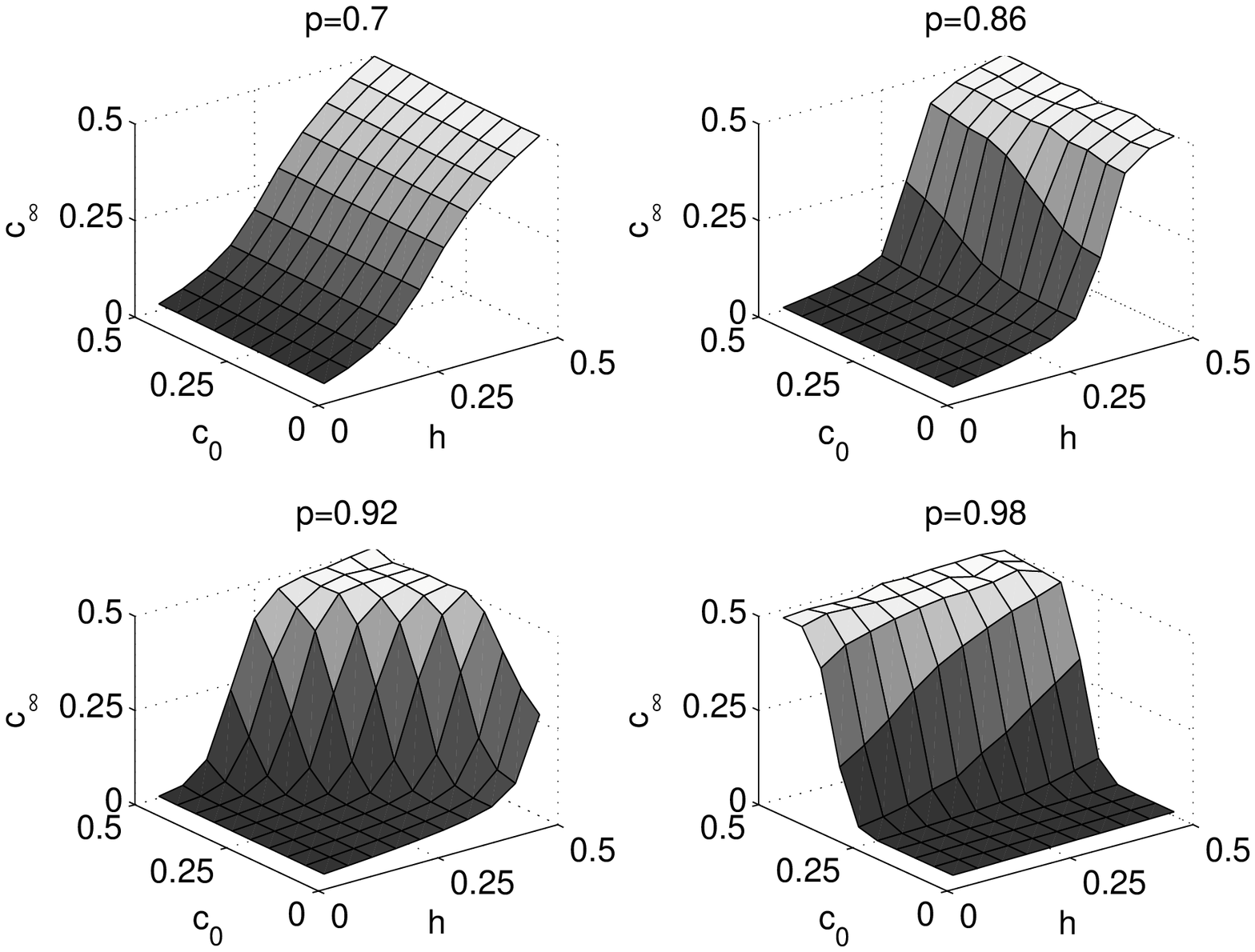}
\caption{MC results for the CAP model and $p\ge 0.7$}
\label{fig:fig_c_inf2_n}
\end{figure}

\begin{figure}[tbp]
\centering
\includegraphics[width=13cm]{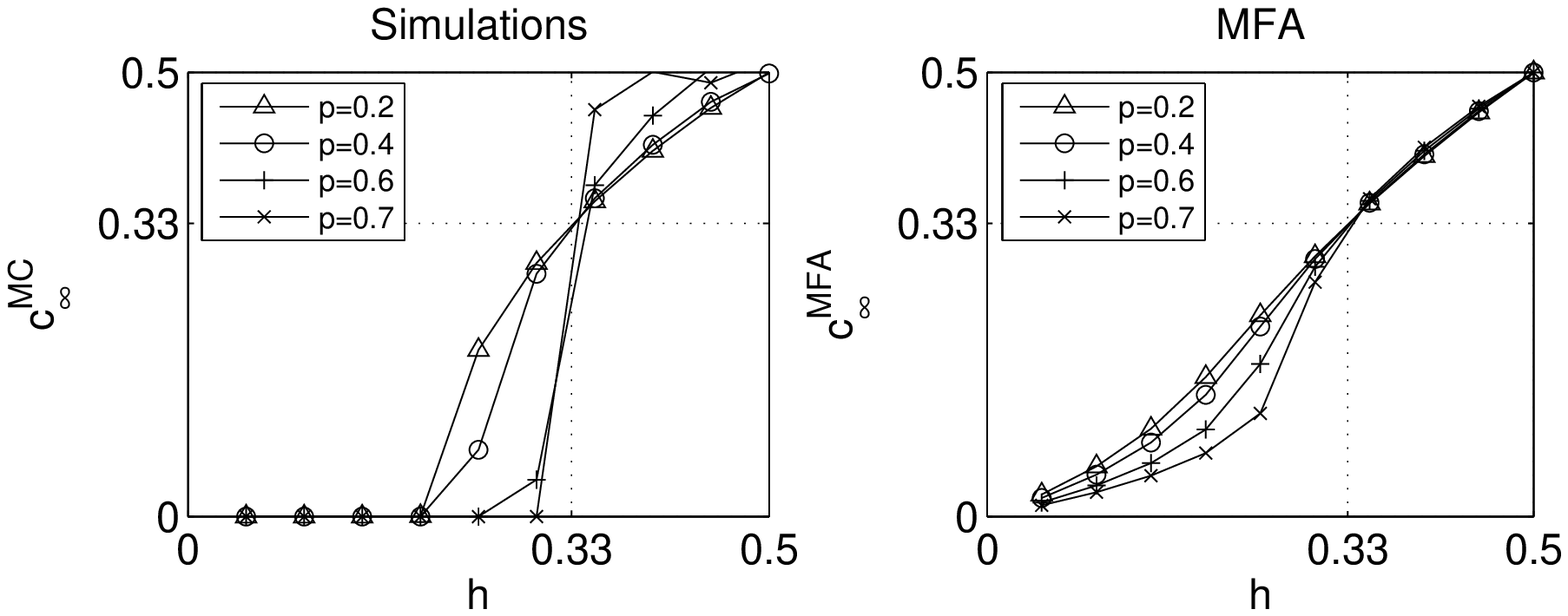}
\caption{MC and MFA results for the CF model and $p\le 0.7$}
\label{fig:fig_c_inf1_s}
\end{figure}

\begin{figure}[tbp]
\centering
\includegraphics[width=12cm]{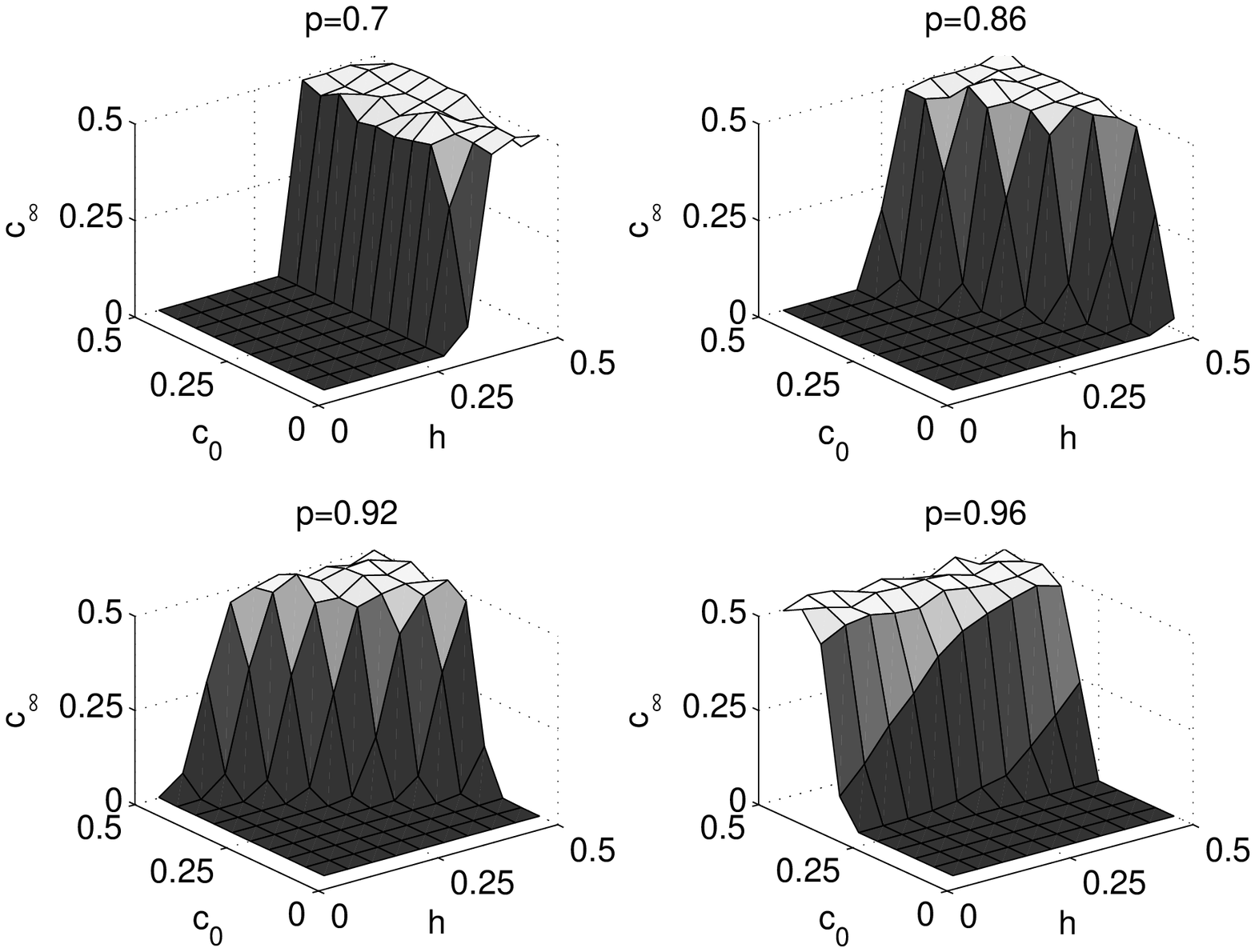}
\caption{MC results for the CF model and $p\ge 0.7$}
\label{fig:fig_c_inf2_s}
\end{figure}

During the simulations, we have observed that after a transient (`termalization') time $t_T$ the concentrations (market shares) reach a final level $c_\infty$ around which they fluctuate. Due to these fluctuations, we define the final value of concentration as the mean value over a time interval  $\Delta T$:
\begin{equation}
c_{\infty}=\frac{1}{\Delta T} \sum_{t=t_T}^{t_T+\Delta T}c(t).
\end{equation}    
For both models (CAP and CF), the final steady state $c_\infty$ was reached in all performed simulations independently of the parameters ($p,h$) and the initial conditions ($c_0$). Moreover, averaging over different $\Delta T$'s (10, 50, 100) did not influence the results. For each set of parameters ($p,h,c_0$) we performed $10^3$ independent Monte Carlo simulations and calculated $c_{\infty}$ by averaging over all samples. The final steady concentration $c_{\infty}$ results are presented in Figures \ref{fig:fig_c_inf1_n}, \ref{fig:fig_c_inf2_n}, \ref{fig:fig_c_inf1_s} and \ref{fig:fig_c_inf2_s}.

For the CAP and CF models two qualitatively different regimes are observed depending on conformity level $p$. For $p \le 0.7$ the final steady state $c_\infty$ does not depend on the initial concentration $c_0$, i.e. it is a function of only two parameters: $c_\infty=f(p,h)$. In the `low conformity' regime ($p \le 0.7$) substantial differences between the studied models can be seen. For the CAP model (see the left panel in Figure \ref{fig:fig_c_inf1_n}), the dependence between final concentration $c_\infty$ and advertising level $h$ is almost linear, but generally $c_\infty=f(p,h)$ is an S-shaped function of $h$. Deviations from linear dependence increase with $p$. %, but $c_\infty=h$ for $p \rightarrow 0$. 
A different behavior is observed for the CF model (see the left panel in Figure \ref{fig:fig_c_inf1_s}). A critical value $h_c$ of the advertising level exists:
\begin{eqnarray}
c_\infty=0 \; \mbox{for} \; h<h_c, \nonumber\\
c_\infty>0 \; \mbox{for} \; h>h_c. 
\end{eqnarray}
In the `high conformity' regime ($p \ge 0.7$), the final steady state $c_\infty$ depends not only on conformity and advertising, but also on initial concentration $c_0$. Nevertheless, in the CAP model the dependence between $c_\infty$ and $h$ is still an S-shaped function and no critical value of $h$ exists (see Figure \ref{fig:fig_c_inf2_n}). This is in sharp contrast to the CF model for which $h_c$ exists (see Figure \ref{fig:fig_c_inf2_s}). For $p\ge 0.7$ the dependence between the final and initial concentration grows with conformity level in both models. This is an understandable result, because for high values of $p$  interactions between individuals dominate over the external field (advertising). On the other hand, for high values of $p$ the dependence between $c_\infty$ and $h$ should decrease. In the limit ($p=1$) only interactions between customers exist and there is no external field, thus no dependence on $h$ is expected. This is confirmed by MC results (see Figures \ref{fig:fig_c_inf2_n} and \ref{fig:fig_c_inf2_s}).

\subsection{Mean Field Results}

\begin{figure}[tbp]
\centering
\includegraphics[width=12cm]{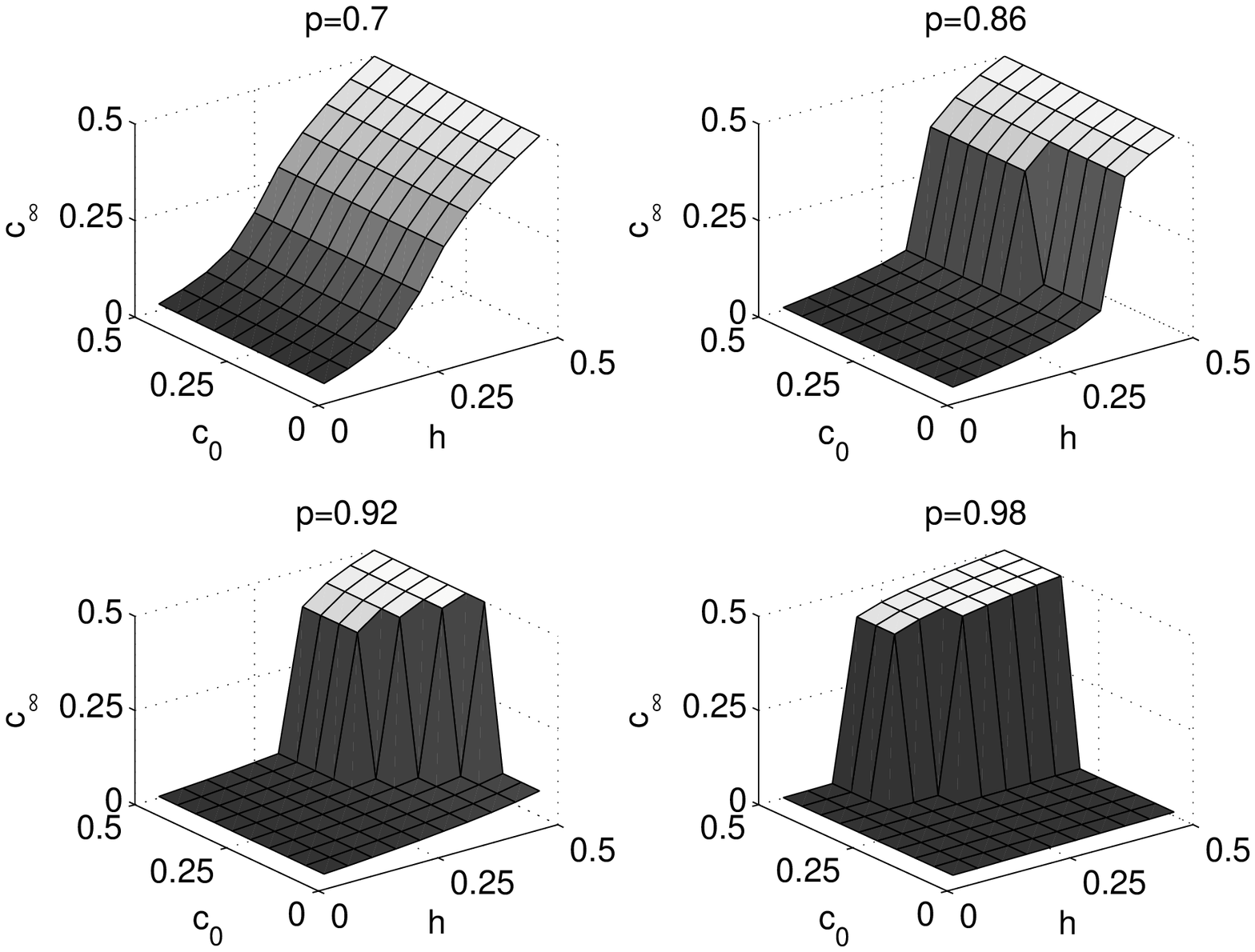}
\caption{MFA results for the CAP model and $p\ge 0.7$}
\label{fig:fig_c_inf2_n_mfa}
\end{figure}

\begin{figure}[tbp]
\centering
\includegraphics[width=12cm]{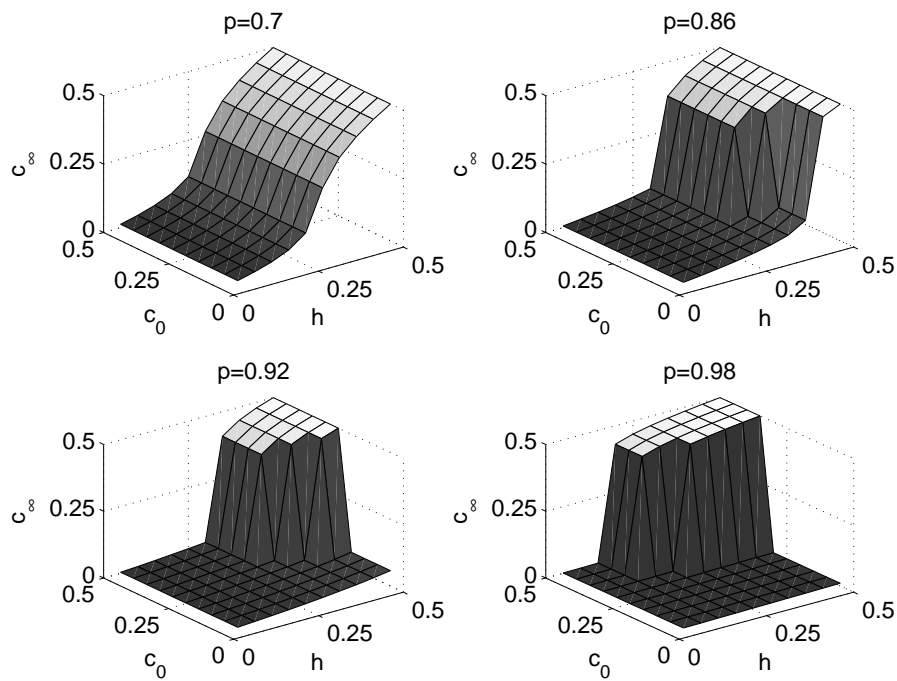}
\caption{MFA results for the CF model and $p\ge 0.7$}
\label{fig:fig_c_inf2_s_mfa}
\end{figure}

In Section \ref{ssec:MFA} we have derived general sets of evolution equations describing how concentrations change in time for both models. These sets of equations can be easily solved numerically. Recall from Section \ref{ssec:PolishMarket}, that we limit the parameter space ($p, h_s, c_s(0)$) to the situation where the initial concentrations of clients and the levels of advertisement of \emph{Plus} and \emph{Era} are equal, i.e. $c_1(0)=c_2(0)\equiv c_0$ and $h_1=h_2\equiv h$, respectively. Then the MFA equations (\ref{eqn:MFA:CAP}) and (\ref{eqn:MFA:CF}) lead to the final fixed point $c_\infty = f(c_0,p,h)$ independently of the parameters ($c_0,p,h$). In agreement with MC simulations, the MFA equations lead to two regimes depending on conformity level $p$:
\begin{itemize}
\item
For $p \le 0.7$ the final steady state $c_\infty$ does not depend on initial concentration $c_0$, i.e. it is function of only two parameters $c_\infty=f(p,h)$, see Figures \ref{fig:fig_c_inf1_n} and \ref{fig:fig_c_inf1_s}.
\item
For $p \ge 0.7$ the final steady state $c_\infty$ depends not only on conformity and advertising level, but also on initial concentration $c_0$. In this regime dependence between the final and initial concentration grows with conformity level $p$. On the other hand, dependence between $c_\infty$ and $h$ decreases and in the limit ($p=1$) there is no dependence on $h$, see Figures \ref{fig:fig_c_inf2_n_mfa} and \ref{fig:fig_c_inf2_s_mfa}. 
\end{itemize}

For the CAP model, MFA and MC results are in a high agreement both in the `low' (see Figure \ref{fig:fig_c_inf1_n}) and `high conformity' regime (Figures \ref{fig:fig_c_inf2_n} and \ref{fig:fig_c_inf2_n_mfa}). The dependence between final concentration $c_\infty$ and advertising level $h$ is an S-shaped function of $h$ and with decreasing $p$ it approaches a linear function: $c_\infty=h$. This similarity indicates a mean field character, i.e. lack of spacial fluctuations, of the CAP model. 
On the contrary, for the CF model MFA results are qualitatively different from those obtained by Monte Carlo simulations, especially in the `low conformity' regime. Most importantly, there is no critical value of $h$ within the Mean Field Approach. 

Monte Carlo results show that, generally, the relationship between market share and advertising level is less straightforward for the CP model due to stronger clusterization caused by a greater role of social validation. Fluctuations within the CP model were able to `produce' a critical phase transition in terms of advertising. Below the critical value $h_c(p)$ (of incumbents' advertising) only the entrant survives. The existence of fluctuations is the reason why MFA for the CP model does not give as compatible results as for the CAP model. It should be noticed that although a critical value of advertising does not exist within the MFA, MFA results still give more complicated dependences between market share and advertising for the CP model than for the CAP model.

\subsection{Fitting to real data}

Up till now we have been investigating general features and differences between CAP and CP models. We have found that generally the CF model is more interesting from a statistical physics point of view. Now we would like to check which model, if any, better describes real data from the Polish telecommunications market. 

The results mentioned below refer to computer simulations for a $100 \times 100$ square lattice with periodic boundary conditions. The initial concentrations $c_s(0), s=1,2,3,$ were set to the market shares of \emph{Plus}, \emph{Era} and \emph{Orange}, respectively, in terms of income for the end of year 2000 (see Section \ref{ssec:PolishMarket}). The levels of advertising $h_s(t), s=1,2,3,$ were set to the respective percentages of total advertising expenditures for the 27 consecutive quarters ($t=1,...,27$), see the bottom right panel in Figure \ref{fig:GRP_Invest}, with one important change. Namely, due to the turmoil around \emph{Era} (see Section \ref{ssec:TurmoilERA}), \emph{Era}'s advertising levels $h_2(t)$ were decreased in the years 2005-2006 ($t=17,...,24$) by 10\%, a value that was found to best represent the inefficiency of advertising in the turmoil period \cite{wlo:08}.
Finally, the time scale had to be set. We have decided to take such a number of time steps $\Delta \tau$  as to allow each client to change operator once every two years (on average): $\frac{100\times100}{8}\times\frac{7.75}{2}=4219$. The rationale for this comes from the fact that in Poland the standard agreement for subscription customers concerns a two year period. To simplify calculations the number 4219 was replaced by 4212, which is divisible by 27.

\begin{figure}[tbp]
\centering
\includegraphics[width=14cm]{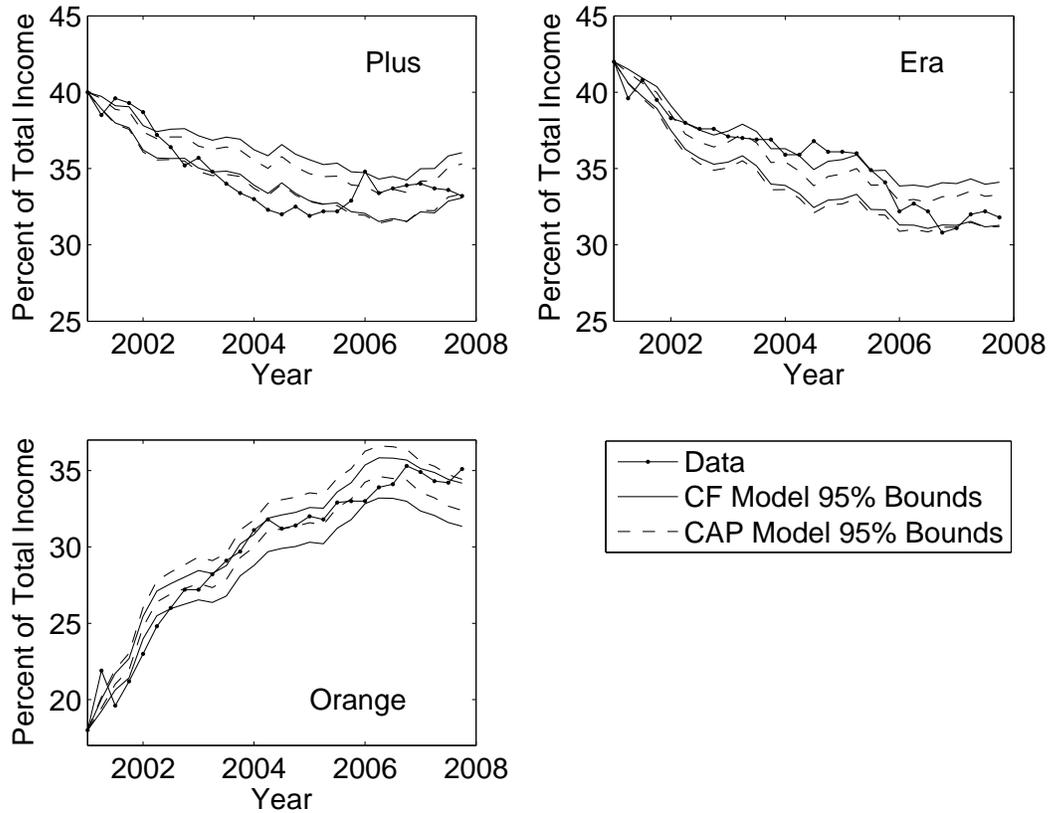}
\caption{Market shares in terms of income for the three operators (\emph{Plus}, \emph{ERA} and \emph{Orange}) and the model generated 95\% bounds for conformity level $p=0.4$. The 95\% bounds are obtained as the 97.5\% quantile line (upper bound) and the 2.5\% quantile line (lower bound) based on 1000 simulated trajectories of each model (CF and CAP).}
\label{fig:LTF}
\end{figure}

Simulation results are presented in Figure \ref{fig:LTF}. The market shares in terms of income for the three operators (\emph{Plus}, \emph{ERA} and \emph{Orange}) and the CF and CAP model generated 95\% bounds are displayed. The 95\% bounds are obtained as the 97.5\% quantile line (upper bound) and the 2.5\% quantile line (lower bound) based on 1000 simulated trajectories of each model. The conformity level was set to $p=0.4$. This choice is not accidental. For both models the best fits to real data were obtained for conformity level $p \in (0.3,0.4)$. Interestingly, this agrees very well with the conformity level found by Asch in his famous social experiment \cite{asc:55}.
Looking at the plots we can conclude that slightly better fits were obtained for the CF model, partly because the bounds were wider. However, we are fitting theoretical models to only one set of data and, hence, we cannot definitely claim that the CF model describes reality better than the CAP model. Obviously, tests with other empirical data sets are needed. 

\section{Conclusions}
\label{sec:Conclusions}

In this paper we have introduced two models of opinion dynamics in an oligopoly market. We have applied them to a situation, where a new entrant challenges two incumbents of the same size. 
Two forces influencing consumer choice were considered: (local) social interactions and (global) advertising. In the Conformity First (CF) model, conformity -- but only in the case of unanimity -- comes in first, then advertising affects the customers. A different idea of incorporating advertising, borrowed from \cite{wol:sta:kul:07}, was used to build the Conformity and Advertising Parallel (CAP) model, where conformity influences individuals parallel to advertising. 

We have studied the general behavior of the models: steady states in terms of market shares for the entire space of parameters, as well as calibrating models to real data. For studying the steady states we have used both the Mean Field Approach and Monte Carlo Simulations. This allowed us to determine the final steady state in terms of market shares $c_\infty$ as a function of initial market shares $c_0$, the level of conformity $p$ and the level of advertising $h$, i.e. $c_\infty=f(c_0,p,h)$. It occurred that both techniques gave very similar results for the CAP model, which indicated the lack of spacial fluctuation within this model. Moreover, the dependence between the final value of market share $c_\infty$ and the level of advertising $h$ was relatively simple and for small values of conformity level $p$ it approached a linear function $c_\infty=h$. On the contrary, for the CP model MFA gave qualitatively different results from MC simulations. Most importantly, there was no critical value of advertising level within the Mean Field Approach, while such a criticality was observed for Monte Carlo simulations. Similar results were obtained earlier for a duopoly market \cite{s-w:wer:03}.

Existence of the critical value is very interesting from the social point of view -- below a certain critical level of $h$ the market approaches a lock-in situation, where one market leader dominates the market and all other brands disappear. 
Groot \cite{gro:06} associates such a market with a situation when consumers tend to copy the behavior of their neighbors (high conformity level) and want the best value for their money, while an open market arises when consumers ignore what their friends buy (low conformity level). In our models the dependence between final market shares and the level of advertising becomes less straightforward with increasing conformity. For low conformity levels advertising becomes the main force of market changes.
Furthermore, Robertson and Gatignon \cite{rob:gat:91} point out that incumbents have an advantage over new entrants, but firms without a responsive defense strategy may forfeit that advantage (this happens in the CF model below the critical advertising level). 

We have also calibrated the models to real data from the Polish telecommunications market. Slightly better fits were obtained for the CF model. However, we are fitting theoretical models to only one set of data and, hence, we cannot definitely claim that the CF model describes reality better than the CAP model. Obviously, more empirical tests with real data are needed. Interestingly, for both models the best fits to real data were obtained for conformity level $p \in (0.3,0.4)$. This agrees very well with the conformity level found by Asch in his famous social experiment \cite{asc:55}. One might say that this agreement is accidental. On the other hand, it might occur that collecting and analyzing data sets in different countries one could determine the `optimal' level of conformity for each country. 

\section*{Acknowledgments}

The authors would like to thank AGB Nielsen Media Research for providing GRP and advertising expenditures data.
Katarzyna Sznajd-Weron gratefully acknowledges the financial support of the Polish Ministry of Science and Higher Education through the scientific grant no. N N202 0194 33.


\section*{References}

\begin{thebibliography}{99}

\bibitem{erd:kea:96} % Decision-making under uncertainty: Capturing dynamic brand choice processes in turbulent consumer goods markets.
Erdem T, Keane MP, 1996 Market. Sci. 15 1

\bibitem{jan:jag:01} % Fashions, habits and changing preferences: simulation of psychological factors affecting market dynamics
Janssen MA, Jager W, 2001 J. Econ. Psychol. 22 745

\bibitem{tur:lee:yin:00} % Customer confusion: the mobile phone market.
Turnbull PW, Leek S and Ying G, 2000 J. Marketing Management 16 143

\bibitem{lee:cha:06} % Consumer confusion in the Thai mobile phone market
Leek S and Chansawatkit S, 2006 J. Consumer Behav. 5 518

\bibitem{bec:mur:93} % A Simple Theory of Advertising as a Good or Bad
Becker GS and Murphy KM, 1993 Q. J. Econ. 108 941

\bibitem{soh:cho:01} % Analysis of advertising lifetime for mobile phone
Sohn SY and Choi H, 2001 Omega 29 473

\bibitem{uke:08}
Urz\c{a}d Komunikacji Elektronicznej, 2008 \emph{Report on the telecommunications market in 2007} (Warsaw: UKE)

\bibitem{cho:lee:chu:01} % Competition in Korean mobile telecommunications market: business strategy and regulatory environment
Choi S-K, Lee M-H and Chung G-H, 2001 Telecommunications Policy 25 125 

\bibitem{dub:hit:man:05} % An Empirical Model of Advertising Dynamics
Dub\'e J-P, Hitsch G J and Manchanda P, 2005 Quant. Market. Econ. 3(2) 107

\bibitem{gro:06} % Influence of social networks and advertisements on sales of goods (ketchup, mayonnaise and curry)
Groot R D, 2006 Physica A 363 446

\bibitem{lev:99}
Levine JM, 1999 Pers. Soc. Psychol. Rev. 3 358

\bibitem{asc:55}
Asch SE, 1955 Scientific American 193 31

\bibitem{mye:06}
Myers DG, 2006 \emph{Social Psychology} (McGraw-Hill, 9th ed.)

\bibitem{s-w:szn:00} % Original Sznajd model
Sznajd-Weron K and Sznajd J, 2000 Int. J. Mod. Phys. C 11 1157

\bibitem{sta:sou:oli:00}
Stauffer D, Sousa AO and De Oliveira M, 2000 Int. J. Mod. Phys. C 11 1239

\bibitem{gal:86}
Galam S, 1986 J. Math. Psychology 30 426

\bibitem{gal:90}
Galam S, 1990 J. Stat. Phys. 61 943

\bibitem{kra:red:03}
Krapivsky PL and Redner S, 2003 Phys. Rev. Lett. 90 238701

\bibitem{s-w:kru:06}
Sznajd-Weron K and Krupa S, 2006 Phys. Rev. E 74 031109

\bibitem{sta:02}
Stauffer D, 2002 Comput. Phys. Commun. 146 93

\bibitem{sch:02}
Schechter B, 2002 New Scientist 175 42

\bibitem{for:sta:05}
Fortunato S and Stauffer D, 2005 in:  Albeverio S, Jentsch V and Kantz H (eds) \emph{Extreme Events
in Nature and Society} (Berlin: Springer)

\bibitem{s-w:05}
Sznajd-Weron K, 2005 Acta Phys. Pol. B 36 2537 

\bibitem{cas:for:lor:07}
Castellano C, Fortunato S and Loreto V, 2007 arXiv:0710.3256v1

\bibitem{sch:03}
Schulze C, 2003 Int. J. Mod. Phys. C 14 95

\bibitem{s-w:wer:03} % How effective is advertising in duopoly markets?
Sznajd-Weron K and Weron R, 2003 Physica A 324 437

\bibitem{can:maz:08} % Mass media influence spreading in social networks with community structure
Candia J and Mazzitello K I, 2008 J. Stat. Mech. P07007

\bibitem{wol:sta:kul:07}
Wo{\l}oszyn M, Stauffer D and Ku{\l}akowski K, 2007 Physica A 378 453

\bibitem{wlo:08}
W{\l}oszczowska M, 2008 \emph{Cola Wars -- the power of advertising in an oligopoly market} (MSc Thesis: Wroc{\l}aw University of Technology; in Polish)

\bibitem{rob:gat:91} % How innovators thwart new entrants into their market.
Robertson TS and Gatignon H, 1991 Planning Rev. 19 4

\end{thebibliography}
\end{document}